\documentclass[prl,reprint]{revtex4-1}
\usepackage{ae}
\usepackage[pdftex]{graphicx}
\usepackage{amsmath}
\usepackage{amssymb}
\usepackage{epsfig, float}
\usepackage{hyperref}
\usepackage{color}

\DeclareMathOperator*{\argmax}{arg\,max}

\newcommand{\vect}[1]{\boldsymbol{#1}}

\begin{document}
\date{\today }

\title{Universality and correlations in individuals wandering \\ through an online extremist space}

\author
{Z. Cao$^1$, M. Zheng$^1$, Y. Vorobyeva$^2$, C. Song$^1$, N.F. Johnson$^1$}

\affiliation{$^{1}$Physics Department, University of Miami, Coral Gables, FL 33146, U.S.A.}

\affiliation{$^2$Dept. of International Studies, University of Miami, Coral Gables, FL 33146, U.S.A.}


\begin{abstract}
The `out of the blue' nature of recent terror attacks and the diversity of apparent  motives, highlight the importance of understanding the online trajectories that individuals follow prior to developing high levels of extremist support. Here we show that the physics of stochastic walks, with and without temporal correlation, provides a unifying description of these online trajectories. Our unique dataset comprising all users of a global social media site, reveals universal characteristics in individuals' online lifetimes. Our accompanying theory generates analytical and numerical solutions that describe the characteristics shown by individuals that go on to develop high levels of extremist support, and those that do not. The existence of these temporal and also many-body correlations suggests that existing physics machinery can be used to quantify and perhaps mitigate the risk of future events. 
\end{abstract}

\maketitle

Following the terror attacks in London, Manchester, Washington D.C. and Paris in 2017, and Orlando, Berlin, Nice and Brussels in 2016, authorities face the fundamental problem of detecting individuals who are currently developing intent in the form of strong support for some extremist entity -- even if they never end up doing anything in the real world. The importance of online connectivity in developing intent \cite{4,5,6,7,8,3,1,9} has been confirmed by case-studies of already convicted terrorists by Gill and others \cite{4,5,6,7,8}. Quantifying this online dynamical development can help move beyond static watch-list identifiers such as ethnic background or immigration status. Heuristically, one might imagine that an individual who enters an online space, wanders through the content available and -- depending in part on what they find on any given day -- feels pulled toward, or pushed away from, a particular extreme ideology. This process of individual fluctuation will be made even more complex by endogenous and exogenous factors in their own lives. Adding to the complication, humans are heterogeneous and hence may enter an online space at different times, spend different amounts of time online, and may end up losing interest and dropping out entirely, continuing in an uncertain state, or developing a high level of support. 

We show here that despite this wide range of possible behaviors, a surprising level of universality arises in the online trajectories of individuals through an extremist space. We provide a stochastic walk model that connects together all individuals, even though they may end up with very different outcomes. Though our focus is on individuals' online dynamics irrespective of whether they later carry out an extremist act or not, subsequent analysis of media reports together with others' postings suggest that a significant number of individuals in our dataset do. Our dataset is assembled using the same methodology as Ref. \cite{scienceus}, and includes the global population of $\sim 350$ million users of the social media outlet VKontakte (www.vk.com) which became the primary online social media source for ISIS propaganda and recruiting during 2015 \cite{scienceus}. Unlike on Facebook where pro-ISIS activity is almost immediately blocked, support on VKontakte develops around online groups (i.e. self-organized communities) which are akin to Facebook groups that support everyday topics such as a sport team. 
\begin{figure}[H]
\begin{center}
\includegraphics[width=0.6\linewidth]{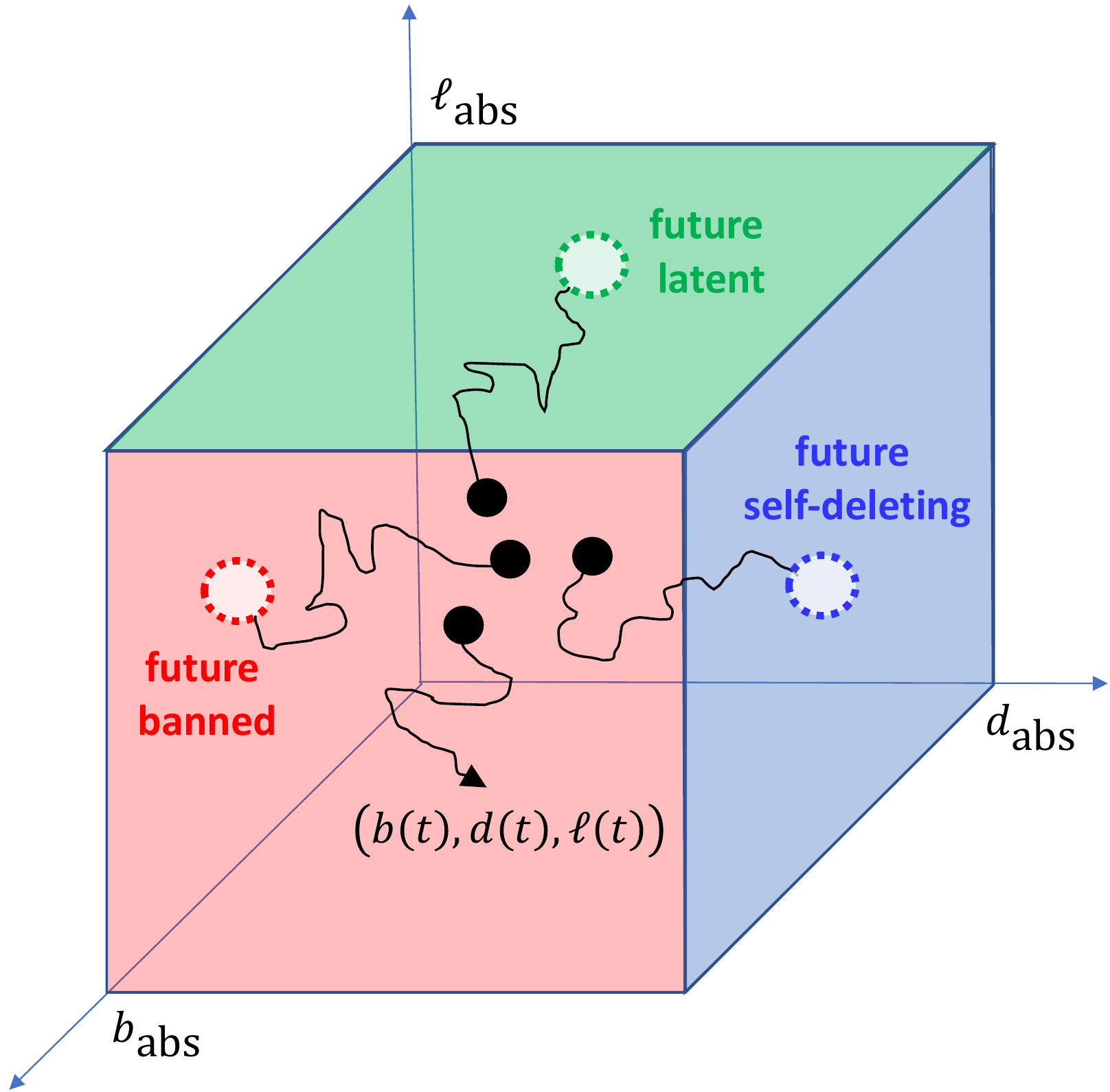}
\end{center}
\caption{(Color online) Schematic of possible individual trajectories in $d-\ell-b$ space bounded by absorbing barriers at $d_{\rm abs},\ell_{\rm abs},b_{\rm abs}$ such that $d(t)<d_{\rm abs}$, $\ell(t)<\ell_{\rm abs}$ and $b(t)<b_{\rm abs}$. These illustrate the three possible outcomes of interest.}
\end{figure}

Given that other forms of extremism ranging from far-left to far-right {\em also} appear through such online groups (e.g. the Washington D.C. shooter \cite{DC} and Maryland attacker \cite{Maryland} were both members of such groups on Facebook), and given that many social media sites allow such community features (i.e. online groups), our results and model should have general applicability. Even the encrypted application Telegram allows users to set up `super-groups' \cite{9}.  All these online groups tend to keep themselves open-source in order to attract new members, hence we were able to record the current membership of pro-ISIS groups at every instant using entirely open-source information. Each individual moving through such an online space can be classified at any time $t$ by what will happen to them {\em in the future}, even though he/she may at time $t$ still be  undecided about supporting the ideology, or may even be moving away from it. Each individual at time $t$ has one of four unique labels: 

\noindent {\em Future banned}: At some future time, he/she will develop and hence express such a high level of extremist support that their account will get banned by moderators.  These individuals would likely be of {\em most} interest to authorities. 

\noindent {\em Future latent}: At some future time, he/she will stop being a member of extremist groups but will not self-delete their account, perhaps reflecting indifference to the extremist ideology. 

\noindent {\em Future self-deleting}: At some future time, he/she self-deletes their own account, perhaps because they are scared of being tracked. 

\noindent {\em Still ongoing}: He/she will remain in development. Their account remains unbanned and they continue joining pro-ISIS groups (and possibly leaving, though group leaving events are rare). 

We focus here on the first three individual types since they provide us with well-defined lifetimes and timelines in terms of the online extremist groups that they join. Banning and self-deleting events are announced on a user's webpage by moderators when they occur. The clock-time lifetimes are $T_{\rm fut-ban}$, $T_{\rm fut-latent}$, and $T_{\rm fut-sdel}$: $T_{\rm fut-ban}$ is the time interval between them first joining a pro-ISIS group and their account being banned; $T_{\rm fut-latent}$ is the time interval between them first joining a pro-ISIS group and them ceasing to be a member of any pro-ISIS group; and $T_{\rm fut-sdel}$ is the time interval between them first joining a pro-ISIS group and them self-deleting their account. The event-time lifetimes $L_{\rm fut-ban}$, $L_{\rm fut-latent}$, and $L_{\rm fut-sdel}$, are given by the total number of groups that they join during the observation period. We model the instant of banning as an individual hitting an absorbing barrier for the first time at $b_{\rm abs}$ in a one-dimensional walk $b(t)$, where $b(t)$ represents the level of extremism (i.e. pro-ISIS support) that an individual expresses. The instant of becoming latent is when an individual hits an absorbing barrier $\ell_{\rm abs}$ during a one-dimensional walk $\ell(t)$, where $\ell(t)$ represents the desire to become latent. The instant of self-deletion is when an individual hits an absorbing barrier at $d_{\rm abs}$ in a one-dimensional walk $d(t)$, where $d(t)$ represents an individual's desire to self-delete.  
Though an obvious over-simplification, such a single scalar parameter has already been adopted in other sociological contexts to mimic aspects of human personality \cite{axelrod,Centola}. 

Each of the $\sim 350$ million VKontakte users undergoes their own walk in the three-dimensional $d$-$\ell$-$b$ space in Fig. 1, characterized by the position vector $(b(t), \ell(t), d(t))$ and with absorbing barriers at $d_{\rm abs},\ell_{\rm abs},b_{\rm abs}$ such that $d(t)<d_{\rm abs}$, $\ell(t)<\ell_{\rm abs}$ and $b(t)<b_{\rm abs}$. We identify 7,707 individuals that eventually hit the barrier along the $b$-axis in Fig. 1 (i.e. future banned individuals); 65,169 that eventually hit the barrier along the $\ell$-axis (i.e. future latent individuals); and 18,905 individuals that eventually hit the barrier along the $d$-axis (i.e. future self-deleting individuals). In principle, the components of the walks along each direction could be coupled, however for simplicity we treat each individual as executing a $1+1+1$-dimensional walk \cite{Stanley}. Hence solving for the lifetime $T$ in a generic single dimension $x(t)$ with a single absorbing barrier at $x_{b}$ solves the problem for each of these dimensions and yields a lifetime distribution for the entire process. Allowing for non-zero drift velocity $u$ towards the respective barrier $x_{b}$, the Fokker-Planck equation for any component $x(t)$ in Fig. 1 becomes:
\begin{equation}
\begin{cases}
&(\frac{\partial}{dt}-D\frac{\partial^2}{\partial x^2} + u \frac{\partial}{\partial x})G(x,t;x_0,t_0)=0\\
&G(x,t;x_0,t_0)=\delta(x-x_0)\delta(t-t_0)\\
&G(b,t;x_0,t_0)=0
\end{cases} 
\end{equation}
where $x \leq x_{b}$, $0 \leq t-t_0 \leq T$. $T$ is the observation period and $D$ is the diffusion coefficient, assumed to be time-independent. For the simulations, we consider the discrete version with unit diffusion speed $\Delta x/\Delta t=1$, with $D=(1-u^2)/2$ and $u=2p-1$ where $p$ is the probability of moving forward at each timestep. The solution is $G(x,t;x_0,t_0)=\Theta(t-t_0)K(x,t;x_0,t_0)$, where the propagator
$
K(x,t;x_0,t_0)=\Phi(x-x')-\exp{[{-u(x_{b}-x)}/{D}]}\Phi[x-(2x_{b}-x')],
$ 
$x'=x_0+ u(t-t_0)$, and $ \Phi(x)= \exp{[-{x^2}/{(4 D t)}]}/\sqrt{4 \pi D t}$. 
To mimic human heterogeneity, we consider a uniformly distributed initial condition at $t=t_0=0$:
\begin{equation}\label{eqn:f1d}
f_{1D}(x, t)= \begin{cases} 
	\delta_{t,0}/x_{m} & x_b-x_{m}\leq x < x_b \\
      0 & {\rm elsewhere}
   \end{cases}
\end{equation}
where $\delta$ is the Kronecker delta and $x_m$ is a normalization constant. $x_{m}=T\Delta x/{\Delta t}$ in the simulations. This is reasonable since individuals located below $x=x_b-x_{m}$ can never reach the boundary and hence can be ignored.
The probability distribution  
$
P_{1D}(x,t)=\sum_{t_0=0}^{T} \int_{x_b-x_{m}}^{x_b}dx_0K(x,t;x_0,t_0) f_{1D}(x_0, t_0),
$ and the total probability $R_{S}(t)=\int_{-\infty}^{x_b}dx {P_{1D}(x,t)}
$ (see Supplemental Material (SM)). The distribution of clock-time lifetimes \begin{equation}
F(t)=\frac{-\frac{dR_S(t)}{dt}}{-\int_{0}^{T}dt\frac{dR_S(t)}{dt}}=\frac{-\frac{dR_S(t)}{dt}}{R_S(0)-R_S(T)}\label{eqn:fist-hitting-time1}
\end{equation}
where 
\begin{equation}
\begin{split}
F(t)&=Z^{-1}\bigg\{u \bigg[\psi\Big(\frac{x_{m}-t u}{2 \sqrt{D t}}\Big)+\psi\left(\frac{u}{2}\sqrt{\frac{t}{D}}\right)\bigg]\\
 +&\sqrt{\frac{4D}{\pi  t}} \bigg[\exp{\left(-\frac{t u^2}{4 D}\right)}-\exp{\left(-\frac{(x_{m}-t u)^2}{4 D t}\right)}\bigg]\bigg \},\label{eqn:clock-time lifetime}
\end{split}
\end{equation}
where $Z$, and $Z_0$ below, are normalizations.
When $u\to 0$: 
\begin{equation}
F(t)|_{u\to0}=Z_0^{-1}\sqrt{\frac{D}{\pi t x_{m}^2}}\bigg[1-\exp{\left(-\frac{x_{m}^2}{4 D t}\right)}\bigg]
\end{equation}
\begin{equation}
F(t)|_{u\to0;t\ll x_m}\sim t^{-1/2}\ .
\end{equation}
Hence our theory predicts an approximate power-law distribution for clock-time lifetimes that are not too large, with a negative scaling exponent of magnitude $1/2$.

\begin{figure}[tbp]
\begin{center}
\includegraphics[width=0.8\linewidth]{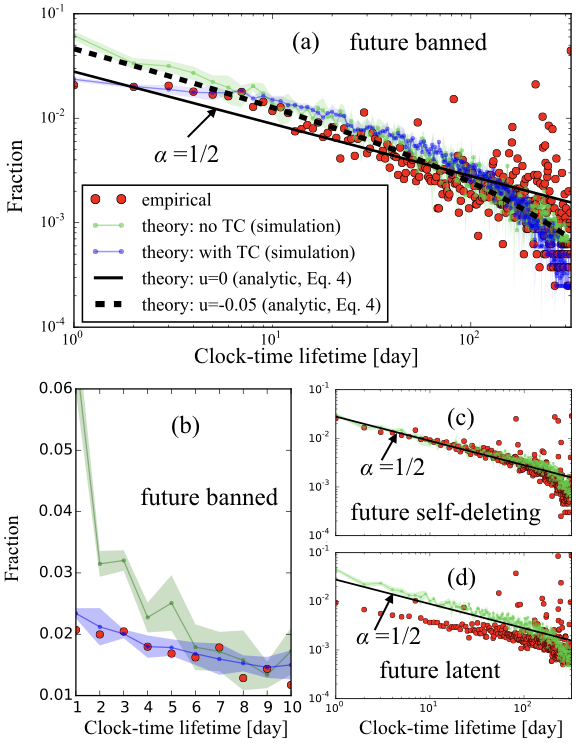}
\end{center}
\caption{(Color online) (a), (c) (d) show the distribution of clock-time lifetimes for individuals with the three types of future outcomes shown in Fig. 1, together with our theoretical predictions. The legend in (a) applies to all. (b) enlarges (a) for short lifetimes. One standard deviation error bands are shown. The TC (i.e. temporal correlation) effect for (c) and (d) is similar to the no-TC case, and hence not shown.}
\label{fig2}
\end{figure}

Figures 2(a),(c),(d) show that despite their very different origins and meanings, all three distributions tend to follow the same analytical $1/2$ power-law for intermediate clock-time lifetimes. Moreover this agreement can be improved by adding a small $u$ in Eq. (4) (e.g. Fig. 2(a)). A full numerical simulation of our model yields even better overall agreement (green curves). Deviations arise at short clock-time lifetimes for future-banned individuals (Figs. 2(a)(b)). However the good agreement can be restored if we add temporal correlations (TC, i.e. memory) to our walk model: with probability $q$, an individual changes his/her $x(t)$ value at time $t$ by adopting the same change that occurred $m$ timesteps earlier. Even the simplest case of $m=1$ shows good agreement (blue curves in Fig. 2). 

\begin{figure}[tbp]
\begin{center}
\includegraphics[width=1.0\linewidth]{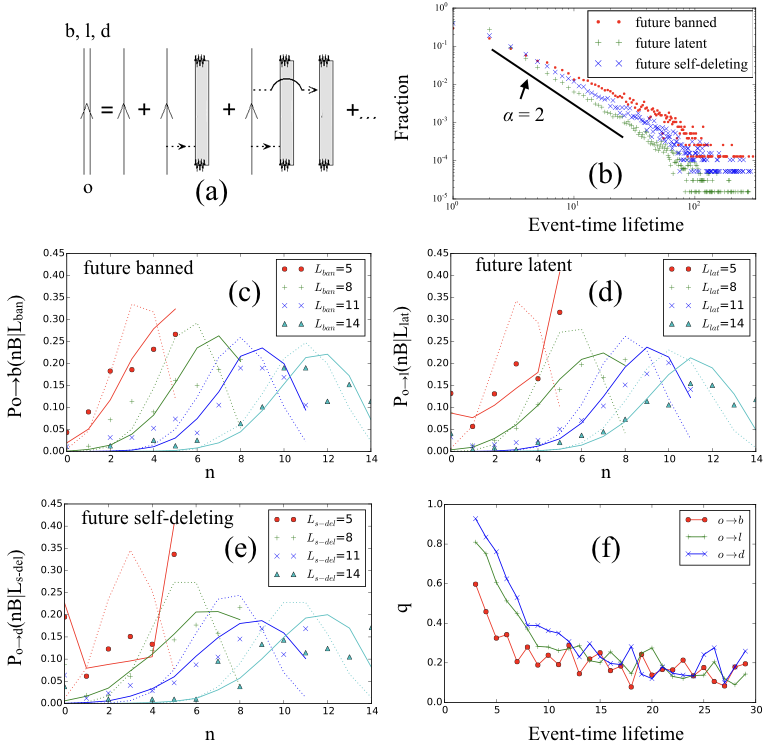}
\end{center}
\caption{(Color online) (a) Exact diagrammatic expansion of the probability that a randomly chosen individual ends up as banned (b), self-deleting (d) or latent (l) after joining $n=0,1,2,\dots$ online groups. $n=0,1,2$ terms are shown explicitly. (b) Terms from (a) evaluated empirically by counting the fraction of individuals who end up as b, d or l after joining exactly $n$ online groups (i.e. event-time lifetime is $n$). (c)-(e) Expansion terms from (a) for a specific event-time lifetime, where here $n$ only counts the number of future-banned groups (B) joined (see text) and hence the maximum $n$ appears bounded from above by the event-time lifetime. Points are empirical results. Solid lines are our temporal correlation (TC, i.e. memory) model results determined using Maximum Likelihood Estimation (MLE). Dashed lines are null model (i.e. binomial distribution). (f) MLE $q$ values in our model. MLE $p$ value $\approx 0.73$ in all cases.}
\end{figure}

Consistent with previous work suggesting that people are highly heterogenous in how long they take to do something \cite{barabasi}, we find that the empirical event-time lifetime can be quite different from the corresponding clock-time lifetime. This motivates us to look at event-time. We represent the probability of an individual ending up in one of the three possible outcomes from Fig. 1, as an expansion (Fig. 3(a)) where each term is the probability that this happens (i.e. lifetime ends) after joining $n$ online groups \cite{10}. Figure 3(b) shows the empirical value of these expansion terms (event-time lifetime is $n$): they follow an approximate power-law distribution with negative scaling exponent $\alpha \approx 2$. While we acknowledge that there are other possible explanations, this is consistent with the notion that someone joining $n$ groups accumulates $n$ potentially distinct narratives and hence may need to resolve $\sim n^2$ potential narrative discords, which in turn suggests that the attractiveness and hence probability of joining $n$ groups will decrease like $n^{-2}$. Figures 3(c)-(e) further unravel individuals' timelines, conditional on the event-time lifetime from (b) and counting as $n$ the number of future-banned groups joined during the lifetime, i.e. $n$ in Figs. (c)-(e) only counts groups who will themselves get banned by moderators. These are of the most interest since by definition they will develop the most extreme content. 

Just as for clock-time, Fig. 3(c)-(e) show that a stochastic walk model without time correlations (TC) provides poor agreement for short event-time lifetimes. Similar to before, we therefore introduce TC (i.e. memory): At each step, with probability $q$ the individual decides to join a group of the same type (i.e. either future-banned or not) as they did in one of their previous $m$ joining events, randomly chosen from $m$. Hence with probability $(1-q)p$, they join a future-banned group, and with probability $(1-q)(1-p)$ they join a non future-banned group. $m$ acts as a memory length, $q$ is the probability of making a decision according to this memory, and $p$ determines individual preference for a specific group type. The model simulation is over 10,000 individuals. $q$ is the dominant parameter in determining the model fit (see SM for details). 

Figure 3(f) shows that $q$, and hence the impact of memory, is most prominent for short event-time lifetimes -- consistent with the conclusion for clock-time lifetimes in Fig. 2. This has an important implication for authorities, since it means that among individuals who will eventually express the most extreme support and hence become banned (i.e. future banned individuals) the ones with shorter lifetimes will exhibit more temporal correlations and hence will exhibit more predictability in their trajectories: and it is precisely these rapidly-developing individuals who likely carry the highest risk of committing future acts.

\begin{figure}[tbp]
\begin{center}
\includegraphics[width=1.0\linewidth]{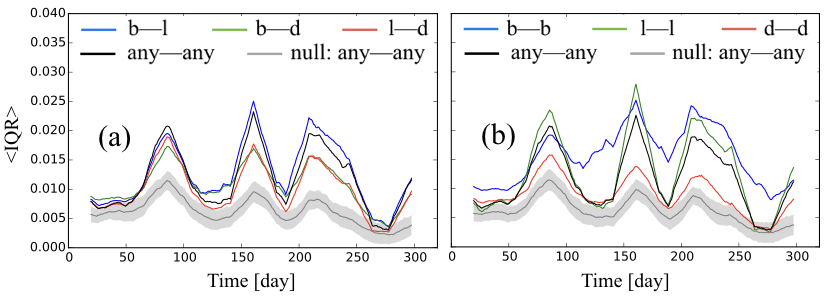}
\end{center}
\caption{(Color online) Evolution of the many-body correlation measure $\langle{\rm IQR}\rangle$ between different sub-populations of individuals. As in Fig. 3(a), b means future banned, l means future latent, d means future self-deleting, and `any' means irrespective of individual type. For example, b--d means between future banned and future latent individuals. One standard deviation error bands are shown around the null model result. In principle, each empirical curve has its own null model, obtained by randomizing the order of the list $\vect{x}_i$ (and $\vect{x}_j$). However these null model results are all very similar, hence we only show the result for the `any --- any' case. See SM for more details.}
\label{fig2}
\end{figure}

Having characterized and quantified the trajectories of individuals, and established the increasing importance of temporal correlations at short lifetimes in both clock-time and event-time, we move to many-body correlations. Though a full theory generalizing the expansion in Fig. 3(a) awaits future development, Fig. 4 shows  the surprising strength and complexity of correlations that evolve over time in the system. Specifically, it shows the average information quality ratio (IQR) \cite{15} of the group joining events (and leaving events, though these are rare) for pairs of individuals of a given type, where
${\rm IQR}(X_i ; X_j) = I(X_i;X_j)/H(X_i,X_j)$. Here $X_i$ and $X_j$ are two  random variables, $I(X_i;X_j)$ is the mutual information of the two random variables, and $H(X_i, X_j)$ is their joint entropy. In our case, $X_i$ and $X_j$ represent the behaviors of  individuals $i$ and $j$ on a given day; therefore IQR becomes an effective measure of the particle-particle correlations. For our dataset, the joint probability density function $P_{X_i, X_j}(x_i, x_j)$ is the probability that two individuals' behavior on any given day is $(x_i, x_j)$, where $x_i$ is measured as the sign of the net change of the number of groups $i$, and therefore $x_i \in \{ -1, 0, 1\}$ (see SM). 

Figure 4 shows that the correlations between trajectories from different sub-populations (Fig. 4(a)) and within the same sub-population (Fig. 4(b)) are all stronger than expected from a null model in which the order of the list $\vect{x}_i$ (and $\vect{x}_j$) is randomized at a given timestep.  This highlights the need to develop an interacting propagator picture for the three individual types. The many-body correlations between future banned users (b--b in Fig. 4(b)) are typically the strongest during the entire period, suggesting that individuals who will go on to develop the most extreme forms of support are the most synchronized. This suggests a new dynamical collective phenomenon by which a relatively small subset of individuals manages to develop coordination within a much larger reservoir of individuals. By contrast, the correlations between future self-deleting individuals are nearly non-existent, suggesting that the movement toward deciding to self-delete is a personal one.

In summary, we have identified, unraveled and quantified the trajectories of individuals wandering through an online extremist space, and found that surprising  statistical universalities exist despite the heterogeneity in individuals' behaviors and their final outcomes.  Our findings establish the increasing importance of temporal correlations at short lifetimes in both clock-time and event-time, which has practical implications for authorities wishing to identify potential high-risk individuals. Our data and results may also help open the path toward a `many-body' theory of human behavior \cite{11,12,13,14} in which single-particle propagators (individuals) successively scatter through dynamical groups that themselves comprise other single-particle propagators, thereby yielding a coupled hierarchy of propagators in a fuller diagrammatic expansion \cite{10}.
 
{\bf Acknowledgments} 
We thank A. Gabriel, A. Kuz, J. Nearing and T. Curtright for initial help with data and discussions. NFJ acknowledges funding under National Science Foundation (NSF) grant CNS1522693 and Air Force (AFOSR) grant FA9550-16-1-0247. The views and conclusions contained herein are solely those of the authors and do not represent official policies or endorsements by any of the entities named in this paper. 


\section*{Supplemental Material (SM)}

\section*{A: Details in the derivation of the clock-time lifespan distribution}

\noindent The explicit form of $P_{1D}(x,t)$:
{\small\begin{equation*}
\begin{split}
P_{1D}(x,t)& = \frac{1}{2x_{m}}\bigg\{\psi\bigg({\frac{x_b+ u t-x}{\sqrt{4 D t}}}\bigg)+ \psi\bigg({\frac{x_{m}+x-x_b- u t}{\sqrt{4 D t}}}\bigg)\\
-&\exp{\bigg[\frac{-u(x_b-x)}{D}\bigg]}\bigg[\psi\bigg({\frac{x_b+x_{m} - u t -x }{\sqrt{4 D t}}}\bigg)\\
+&\psi\bigg({\frac{x+ u t -x_b}{\sqrt{4 D t}}}\bigg)\bigg]\bigg\}
\end{split}
\end{equation*}}
where $\psi(x)$ is the error function. 

\noindent The explicit form of $R_{S}(t)$:
{\small\begin{equation*}
\begin{split}
R_{S}(t)&=\frac{1}{2 x_{m} u}\bigg\{ \left(D-x_{m} u+t u^2\right) \Psi \left(\frac{x_{m}-t u}{2 \sqrt{D t}}\right)\\
& - D \exp{\left(\frac{x_{m} u}{D}\right)} \Psi \left(\frac{x_{m}+t u}{2 \sqrt{D t}}\right)\\
& +2 u \sqrt{\frac{D t}{\pi}} \bigg[\exp{\left(-\frac{(x_{m}-t u)^2}{4 D t}\right)}- \exp{\left(-\frac{t u^2}{4 D}\right)}\bigg]\\
& -\left(2 D+t u^2\right) \psi \left(\frac{u}{2} \sqrt{\frac{t}{D}}\right)+2  x_{m} u- t u^2\bigg\}
\end{split}
\end{equation*}}
where $\Psi(x)$ is the complementary error function. 

\noindent The explicit form of $Z$:
{\small\begin{equation*}
\begin{split}
Z=&\left(\frac{2 D}{u}+T u\right) \psi\left(\frac{1}{2} u \sqrt{\frac{T}{D}}\right)+\frac{D}{u}e^{\frac{x_{m} u}{D}} \Psi\left(\frac{x_{m}+T u}{2 \sqrt{D T}}\right)\\
&+\left(x_{m}-T u -\frac{D}{u}\right) \Psi\left(\frac{x_{m}-T u}{2 \sqrt{D T}}\right)+T u\\
&+2 \sqrt{\frac{D T}{\pi }} \bigg[e^{-\frac{T u^2}{4 D}}-e^{-\frac{(x_{m}-T u)^2}{4 D T}}\bigg]
\end{split}
\end{equation*}} 

\noindent The explicit form of $Z_0$:
{\small\begin{equation*}
Z_0 = 1+ \sqrt{\frac{4D T}{\pi x_m^2}}\bigg[1-\exp\left( - \frac{x_m^2}{4 D T}\right)\bigg]-\psi\left(\frac{x_m}{\sqrt{4 D T}}\right)
\end{equation*}}.

\section*{B: Details of Figs. 3(c)-(f)}
Stochastic simulations show that increasing $m$ would strengthen the memory effect significantly only when $q$ is sufficiently large (e.g., above $\sim 0.7$). Hence for most values of $q$, the profile of the distribution $P_{o \rightarrow b}(n\textrm{B}|L_{ban})$ is primarily determined by $q$. Hence we let $m=1$ for simplicity, and estimate $q$ and $p$ for each value of $L_{ban}$ from the empirical data using maximum likelihood estimation (MLE), and perform a model fit to the empirical results by simulation. For the memory model with $m=1$, the likelihood for an individual $i$ to have a path $\mathbb{S}_i=\{S_{i}[t]|S_{i}[t]\in\{0, 1\}, t=0,1,2,...,L-1\}$ ($S_{i}[t]=0$ corresponds to joining a future-banned group, and $S_{i}[t]=1$ corresponds to joining a non future-banned group) is given by
$
\mathcal{L}_{i}= \prod_{t=0}^{L-2} \{p + q - S_i[t+1] q + 2  S_i[t+1] S_i[t] q - p q + S_i[t] (1 - 2 p - 2 q + 2 p q)
$
Therefore, $p$ and $q$ are given by
$
\argmax_{(q,p)} \bar{\mathcal{L}}\equiv\frac{1}{N}\sum_{i=1}^{N} \mathcal{L}_{i} := \{(q,\ p)| 0\leq q,\ p \leq 1\}
$.
When doing the simulation for a given $L_{ban}$, we do a separate stochastic simulation of 10,000 individuals; and since there is no history in the first step, we randomly assign the initial memory for each individual (the same for other two types of individuals).

\section*{C: Details of  $\langle{\rm IQR}\rangle$}
Without loss of generality, we here show the calculation of the average IQR (i.e. $\langle{\rm IQR}\rangle$) between the future banned and the future self-deleting users (denoted as b---d) in Fig. 4(a).  First, we pick an individual $i$ from the sub-population of future banned individuals, and an individual $j$ from the sub-population of future self-deleting individuals, and we perform the statistics to obtain $P_{X_i}(x_i)$, $P_{X_j}(x_j)$, as well as $P_{X_i, X_j}(x_i, x_j)$ (see the main text for the definitions). For the values of day $t$, the statistics are done from the $(t-10)$'th day  to the $(t+10)$'th day (i.e. a moving window of size 20 days). In order to reduce noise we smoothened the curve by averaging over every 10 days.
Therefore for individual $i$, the values of $x_i$ form a temporally ordered list $\vect{x}_i$. Hence, the mutual information is given by
$
I(X_i; X_j) = \sum_{x_i, x_j} P_{X_i, X_j}(x_i, x_j)\log_{2}\bigg[ \frac{P_{X_i, X_j}(x_i, x_j)}{P_{X_i}(x_i)P_{X_j}(x_j)}\bigg],
$ 
the joint entropy is given by
$
H(X_i; X_j) = - \sum_{x_i, x_j} P_{X_i, X_j}(x_i, x_j)\log_{2}[ P_{X_i, X_j}(x_i, x_j)],
$ 
and therefore, the IQR is given by
$
IQR(X_i; X_j)=I(X_i; X_j)/H(X_i; X_j). 
$
We ignored user pairs whose joint entropy is zero since they represent mostly trivial cases in which no joining/leaving events occurred. Finally, we average over all combinations of the user pairs to obtain the average IQR. Since our dataset is so large, we sampled 2000 users for each individual type 10 times in order to obtain the mean values and their standard deviations.

\end{document}